\documentstyle[epsf,aps,floats,amssymb,preprint]{revtex}
\tightenlines
\newcommand{\chisq}{\mbox{$\chi^{2}$}}
\newcommand{\pbarp}{\mbox{$\bar{p}p$}}
\newcommand{\Et}{\mbox{$E_{T}$}}
\input psfig
\input epsf

\title{Uncertainties in the measurements of the Inclusive Jet Cross Section at the Tevatron}

\author{I.A. Bertram \thanks{Submitted to the {\it VII International Workshop on
         Deep Inelastic Scattering and QCD - DIS99,}
          \hfill\break
          April 19 -- 23, 1999, Zeuthen, Germany.}
	\thanks{E-Mail: bertram@fnal.gov}}
\address{Lancaster University, Lancaster, United Kingdom}
\begin{document}

\maketitle

\begin{abstract}
The systematic uncertainties of the measurements of the Inclusive Jet
Cross Section at the Tevatron and their effect on the \chisq\
comparisons between data and theoretical predictions are discussed.
\end{abstract}

\section{Introduction}

 The inclusive jet cross section in \pbarp\ collisions has recently
 been measured by the CDF~\cite{CDF} and D\O~\cite{d0_inc}
 collaborations. These measurements are compared with NLO perturbative
 QCD predictions~\cite{eks,jetrad}.  These experimental measurements
 have uncertainties that are smaller than the the uncertainties of the
 theoretical predictions,
 $\sim$30$\%$~\cite{inc_jet_theory_uncertainties}.\\

 The CDF measurement of the inclusive jet cross section showed an
 excess of jet production at high transverse energy (\Et ) which could
 be caused by new physics such as quark compositeness, inaccuracies in
 the parton distribution functions, or inadequacies in the NLO QCD
 predictions. The theoretical predictions are in good agreement with
 the D\O\ measurement. Both experimental measurements are also in
 agreement~\cite{jets_review}. Our ability to compare quantitatively
 the theoretical predictions and the measurements depends on a
 thorough understanding of the systematic uncertainties.

\section{Systematic Uncertainties}

\begin{figure}[htb]
\begin{minipage}[t]{3.05in}
{\centerline{\psfig{figure=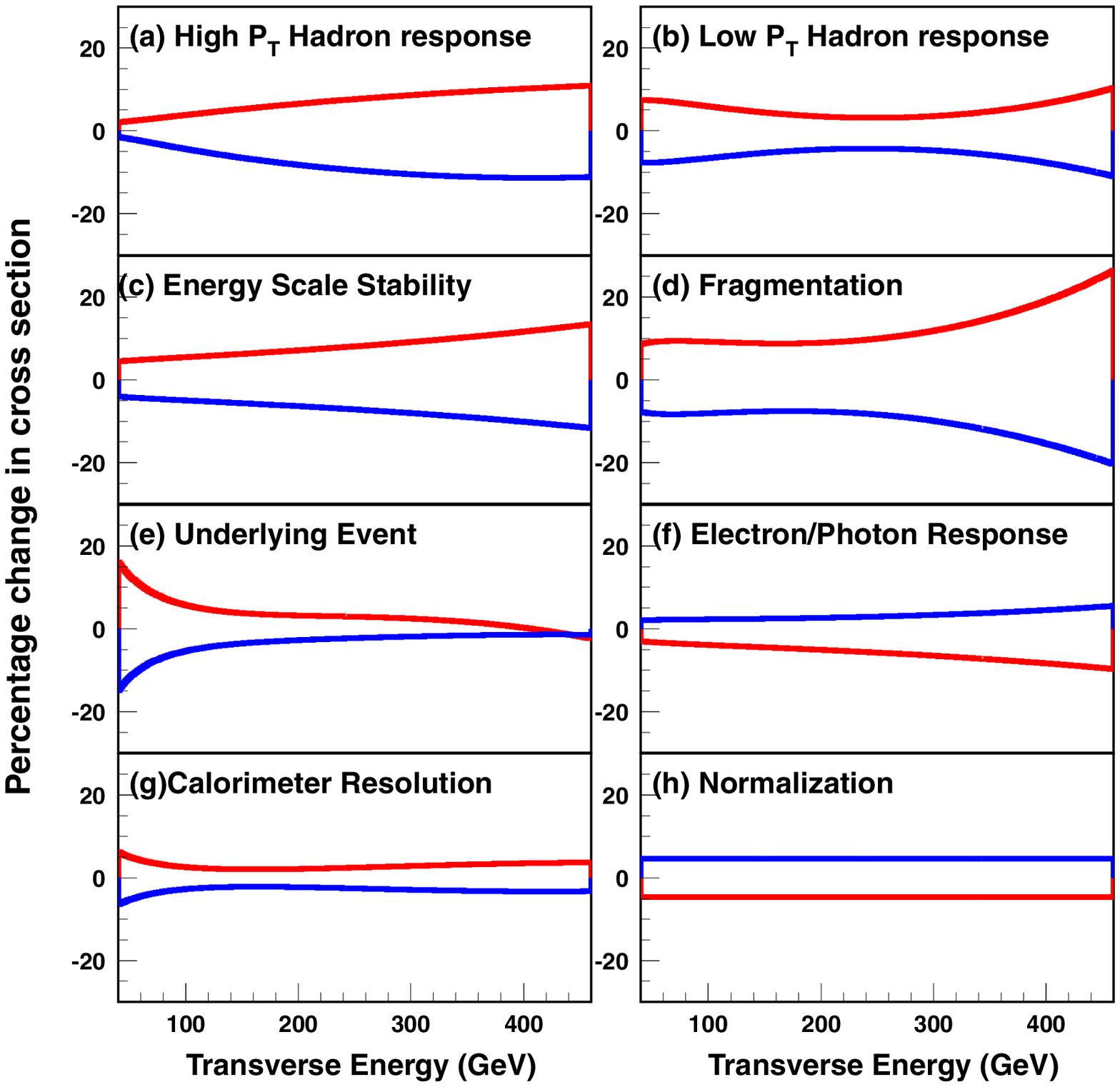,width=3.05in}}
\caption{The systematic uncertainties of the inclusive jet cross section as measured by CDF.}
\label{FIG:cdf_uncertainties}}
\end{minipage}
\hspace*{2mm}
\begin{minipage}[t]{3.05in}
{\centerline{\psfig{figure=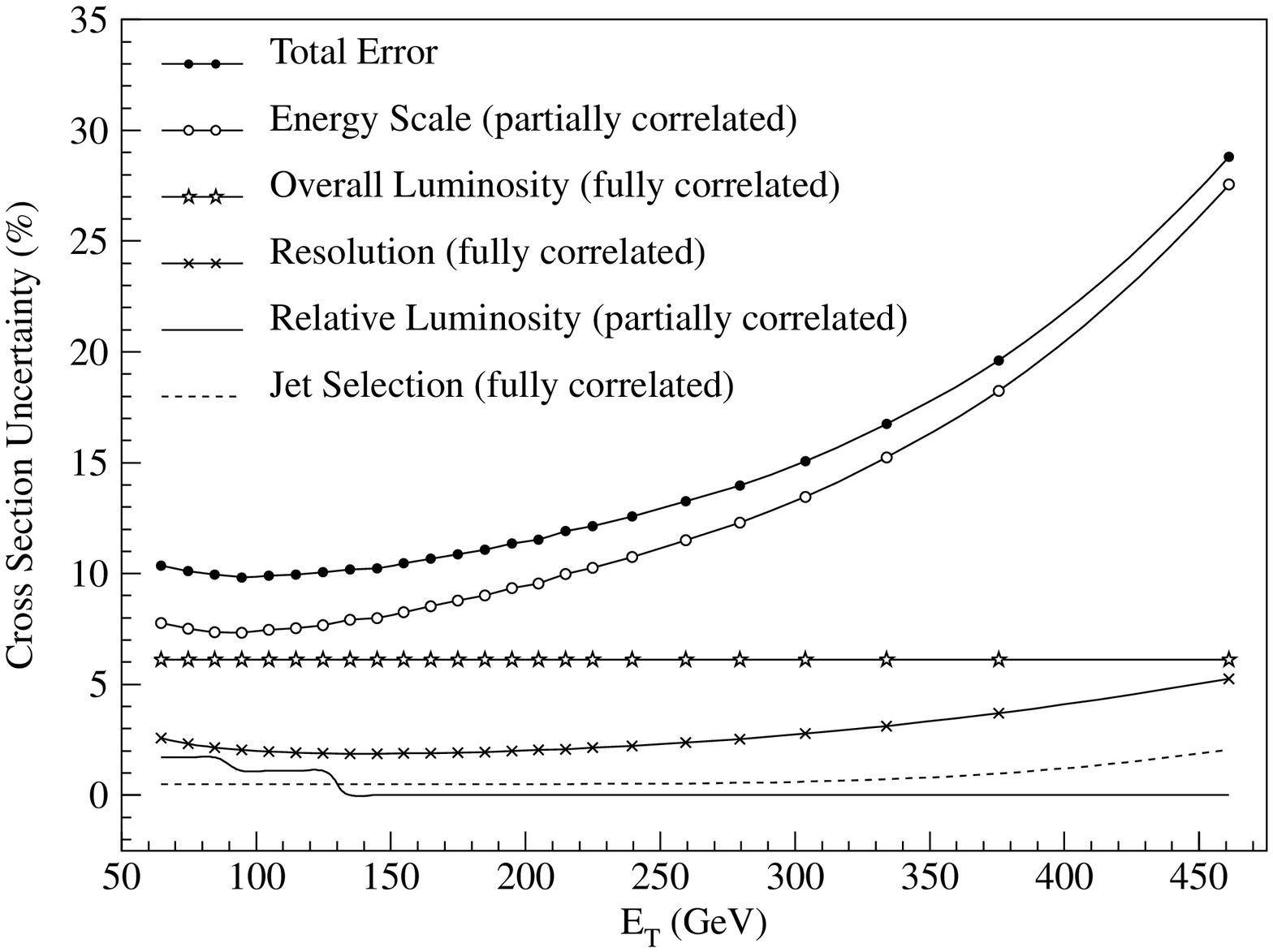,width=3.05in}}
\caption{The systematic uncertainties of the inclusive jet cross section as 
 measured by D\O .}
\label{FIG:d0_uncertainties}}
\end{minipage}

\end{figure}

 The major components of the systematic uncertainties of the CDF
 measurement are depicted in Fig.~\ref{FIG:cdf_uncertainties}. The
 dominant uncertainties are due to the jet energy scale correction,
 the resolution unsmearing, and the integrated luminosity. The
 uncertainties are divided up into different components (see
 Fig.~\ref{FIG:cdf_uncertainties}). Each component is assumed to be
 $100 \%$ correlated as a function of \Et\ and independent of all
 other components.\\
 
 Similarily, the five largest uncertainties in the D\O\ measurement
 are depicted in Fig.~\ref{FIG:d0_uncertainties}. The uncertainties
 are either $100 \%$ correlated , partially correlated (correlation
 lies between $-100\%$ to $100\%$), or uncorrelated as a function of
 \Et .\\

%\begin{figure}[htb]
%\vspace{9pt}
%\end{figure} 

 In general the uncertainties in the jet energy scale, jet energy
 resolutions, the luminosity, {\rm etc.}, are assumed to be Gaussian.
 Hence, the uncertainties of the cross section are asymmetric,
 i.e. the positive and negative errors on the cross section are
 different. This is a direct result of the steeply falling inclusive
 jet cross section.\\

 The assumption of Gaussian uncertainties is not always a valid
 one. One of the major sources of uncertainty in the inclusive jet
 cross section is the integrated luminosity. The two experiments base
 their luminosity calculations on different measurements of the total
 \pbarp\ cross section. CDF uses its own measurement~\cite{lumin_cdf}
 while D\O\ uses a world average cross section~\cite{luminosity_1800}
 based on the CDF~\cite{lumin_cdf} and E710~\cite{lumin_e710}
 measurements. This leads to a $7\%$ difference between the
 luminosities quoted by the two experiments. The assumption that the
 uncertainty due to the luminosity is Gaussian in nature is probably
 incorrect.

 \section{Quantitative Comparisons}

 D\O\ has made quantitative comparisons~\cite{d0_inc} between
 theoretical predictions and their measurement base on a \chisq\
 test. The \chisq\ is given by\begin{equation}
\chisq = \sum_{i,j} \delta_{i} V_{ij}^{-1} \delta_{j}
\end{equation}
 where $\displaystyle{\delta_{i}}$ is the difference between the data
 and theory for a given \Et\ bin, and $\displaystyle{V_{ij}}$ is
 element $i,j$ of the covariance matrix:
\begin{equation}
V_{ij} = \rho_{ij} \cdot \Delta \sigma_{i} \cdot \Delta \sigma_{j}.
\end{equation} 
 where $\displaystyle{\Delta \sigma}$ is the sum of the systematic
 uncertainty and the statistical error added in quadrature if $i=j$
 and the systematic uncertainty if $i \neq j$.
 $\displaystyle{\rho_{ij}}$ is the correlation between the
 uncertainties of two \Et\ bins.\\

 The construction of the covariance matrix requires that the
 uncertainties follow a Gaussian distribution. Hence using \chisq\ to
 determine the probability that a theoretical prediction agrees with a
 measurement does not take advantage of all the information available.
 Additionally it does not take into account boundary conditions (for
 example you cannot fluctuate a cross section more than $100\%$ below
 its value).\\

 Parton distribution fits~\cite{cteq} are now using the measurements
 of the inclusive jet cross section to constrain the gluon
 distributions. If the information available in the inclusive jet
 measurements is to be used to best effect then new methods must be
 developed to calculate the probability that a theoretical prediction
 agrees with the data.

 \section{Conclusion}

 The treatment of the systematic uncertainties of the inclusive jet
 cross section in \pbarp\ collisions have been discussed. The \chisq\
 values presented in~\cite{d0_inc} are based on approximations of the
 uncertainties and do not use all of the information available.\\
 
 I thank my colleagues on the D\O\ and CDF experiments for their
 helpful comments, suggestions and discussions.\\

\end{document}